\documentclass[a4paper]{jpconf}
\usepackage{graphicx}
\begin{document}

\title{The KATRIN sensitivity to the neutrino mass and to 
right-handed currents in beta decay}

\author{J Bonn$^1$, K Eitel$^2$, F Gl\"uck$^{3,4}$, 
D Sevilla-Sanchez$^1$ and N Titov$^{5,6}$}

\address{$^1$  Johannes Gutenberg-Universit\"at Mainz, Institut f\"ur Physik, Germany}

\address{$^2$ Forschungszentrum Karlsruhe, Institut f\"ur Kernphysik, Germany}

\address{$^3$ Universit\"at Karlsruhe (TH), Institut f\"ur Experimentelle 
Kernphysik, Germany}

\address{$^4$ Research Institute for Nuclear and Particle Physics,
Theory Dep., Budapest, Hungary}

\address{$^5$ Westf\"alische Wilhelms-Universit\"at M\"unster, Institut f\"ur Kernphysik, Germany} 

\address{$^6$ Institute for Nuclear Research, Troitsk, Russia}

\ead{jbonn@uni-mainz.de, Klaus.Eitel@ik.fzk.de, Ferenc.Glueck@ik.fzk.de, Daniel.Sevilla@mpi-hd.mpg.de, titov@inr.ru}

\begin{abstract}

The aim of the KArlsruhe TRItium Neutrino experiment KATRIN is the determination of the absolute neutrino mass scale down to 0.2 eV, with essentially smaller model dependence than from cosmology and neutrinoless double beta decay.  For this purpose, the integral electron energy spectrum is measured close to the endpoint of molecular tritium beta decay. The endpoint, together with the neutrino 
mass, should be fitted from the KATRIN data as a free parameter.
The right-handed couplings change the electron energy spectrum close to the endpoint, therefore they have some effect also to the precise neutrino mass determination. The statistical calculations show that, using the endpoint as a free parameter, the unaccounted right-handed couplings constrained by many beta decay experiments can change the fitted neutrino
mass value, relative to the true neutrino mass, by not larger than about 5-10 \%.
Using, incorrectly, the endpoint as a fixed input parameter, the above change of the neutrino mass 
can be much larger, order of 100 \%, and for some cases it can happen that for large true neutrino mass value the fitted neutrino mass squared is negative. Publications using fixed endpoint and 
presenting large right-handed coupling effects to the neutrino mass determination are not relevant for the KATRIN experiment.

\end{abstract}

\section{Neutrino mass determination and the endpoint}
\medskip

In the KATRIN experiment the absolute neutrino mass is determined by the measurement of the integral energy spectrum of the electrons coming from beta decay of tritium molecules. The electrons are guided from the tritium source to the detector by magnetic field. Between the source and the detector a large negative potential (-18.6 kV) is applied at the main spectrometer, 
with the aim that only those electrons can reach the detector that have a decay kinetic energy above the value corresponding to this
potential. The transversal energy component (relative to magnetic field) of the electrons is converted into longitudinal energy by using the inverse magnetic mirror effect. Thus it is possible to measure the integral electron energy spectrum simultanously with high statistics and with high precision. For further information about the KATRIN experiment see Ref.  \cite{DesignReport2004}.

The differential electron energy spectrum can be written (in a first approximation, close to the endpoint) as
\begin{equation}
w_{diff}(E)= E_{\nu } \sqrt{E_{\nu }^2-m_\nu^2},
\end{equation}
where $E$ is the relativistic total electron energy, $E_{\nu }=E_0-E$ and $m_\nu$ denote the 
neutrino energy and mass, and $E_0$ is the nominal endpoint (maximum of
$E$, if the neutrino mass is zero). There are several theoretical modifications to this simplified spectrum, the most important of them is due to the 
recoil molecular ion final state distribution
(see Ref. \cite{Doss} for a recent calculation). 
Degenerate neutrino masses are assumed
(the KATRIN experiment is able to find a non-zero neutrino mass only above 0.2 eV).

\begin{figure}[h]
\raisebox{-1cm}{\includegraphics[width=0.45\textwidth]{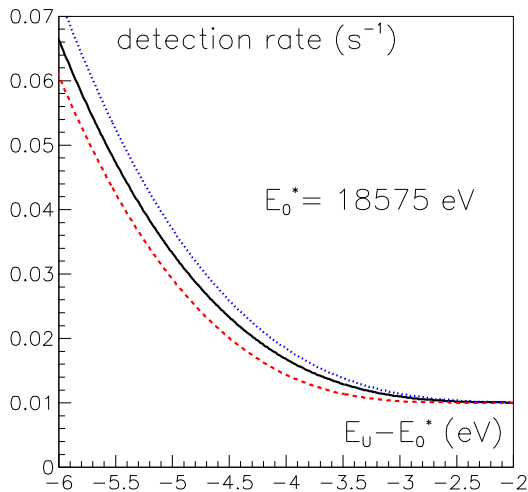}}
\hspace{0.02\textwidth}
\begin{minipage}[b]{0.42\textwidth}\caption{\label{fig1} Expected detection rate of the KATRIN experiment as function of the minimal detected electron energy
$E_U$, for different neutrino mass and endpoint values. Full (black) curve: $m_\nu=0, \; E_0=E_0^*$; dashed (red) curve: $m_\nu=1\;{\rm eV} , \; E_0=E_0^*$; dotted (blue) curve: $m_\nu=0, \; E_0=E_0^*$+{\rm 0.15 eV}.
The new KATRIN design parameters of Ref. \cite{DesignReport2004}
together with $0.01\; s^{-1}$ background rate have been employed.}
\end{minipage}
\end{figure}

The KATRIN experiment measures the integral energy spectrum, therefore
one has to multiply the differential spectrum by the response function of the spectrometer (see Ref. \cite{DesignReport2004} for details), and to integrate from
the minimal electron energy $E_U=e|U_A-U_S|$, where $U_A$ and $U_S$ denote the electric potential in the middle of the main spectrometer (analyzing plane) and in the tritium source, respectively.
The expected absolute detection rate of the KATRIN experiment can be seen in Fig. \ref{fig1} for different neutrino mass and endpoint values. The most sensitive
region for the neutrino mass determination is around $E_U-E_0^*\approx -5$ eV, where the signal is twice
as large as the background (Ref. \cite{Otten}).
It is clear from the figure that there is a positive correlation between the neutrino mass and the endpoint: a larger fixed endpoint value results in a larger fitted neutrino mass value.

In the KATRIN experiment (like in several earlier neutrino mass experiments) the endpoint is a free parameter, to be determined from the
KATRIN spectrum data. Nevertheless, let us assume for a moment that the endpoint is a fixed input parameter. Then a $\Delta E_0$ error of the
endpoint results in a $\Delta m_\nu^2\; ({\rm eV^2})\approx 7 \Delta E_0\; ({\rm eV})$ error for the neutrino mass squared (using the last 20 eV
of the spectrum for the data analysis). From the triton-He3 nuclear mass differences one has at present a $\Delta E_0=1.2$ eV error for the endpoint \cite{Nagy}. In addition, it is difficult to determine
the absolute potential values with a precision better than 100 mV.
On the other hand, the KATRIN experiment aims to measure the 
neutrino mass squared with an accuracy of
$\sigma(m_\nu^2)=0.025\; {\rm eV}^2$. To obtain this precision, the accuracy of the endpoint value (as fixed parameter) should be at least
4 meV. 
Therefore, it is obvious: 
{\bf for the data analysis of the KATRIN experiment the endpoint cannot be used as an external fixed input parameter; it should be used necessarily as a free parameter, determined from the KATRIN data.
Analyses assuming the endpoint as a fixed parameter are not relevant
for the KATRIN experiment.}

\section{Right-handed couplings and the electron energy spectrum}
\medskip

In the presence of right-handed weak couplings the differential electron
spectrum is changed to the following form:
\begin{equation}
\label{wdiff}
w_{diff}(E)=E_{\nu} \sqrt{E_{\nu }^2-m_\nu^2} \left(1+b'\frac{m_\nu}
 {E_{\nu}}\right).
\end{equation}
This formula is valid close to the endpoint. A similar change of the
electron spectrum is due to the Fierz parameter $b$.
The parameter $b'$ is a linear combination of the right-handed 
vector ($R_V$), axial-vector ($R_A$), scalar ($R_S$) and tensor 
($R_T$)couplings:
\begin{equation}
\label{bprimed}
 b'\approx -2 \frac{
 \Re e(L_V R_V^*+L_V R_S^*)|M_F|^2+ \Re e(L_A R_A^*+L_A
 R_T^*)|M_{GT}|^2}
 {|L_V|^2 |M_F|^2+|L_A|^2 |M_{GT}|^2}
\end{equation}
(only the dominant terms are shown in this formula, which is in agreement with Ref. \cite{Enz}). The left-handed
$L_j$ and right-handed $R_j$ couplings have the following simple
relations with the widely used couplings $C_j$ and $C_j'$ introduced by Lee and Yang in Ref. \cite{LeeYang}:
$C_j= \left(L_j+R_j\right)/\sqrt{2}$, 
$C_j'= \left(L_j-R_j\right)/\sqrt{2}$. As it is explained in Ref.
\cite{Gluck95}, there are several advantages using the couplings
$L_j$ and  $R_j$. In the Standard Model only the 
left-handed vector and axial-vector couplings $L_V$ and $L_A$
are non-zero. 

There are many experimental observables (like beta asymmetry, neutrino-electron correlation, beta polarization etc.) that
provide constraints for the couplings $R_j$. Unfortunately, 
these observables are quadratic in the $R_j$ couplings
(with zero neutrino mass the right-handed couplings have
no interference with the dominant left-handed couplings),
therefore the 95 \% confidence limits  are not too small:
$|R_V|<0.08,$ $|R_A|<0.10,$  $|R_S|<0.07,$ $|R_T|<0.10$
(see the recent overview in Ref. \cite{Severijns}; the $L_V=1$ normalization is used here). The signs of the couplings $R_j$
are not known; in order to obtain a conservative limit for $b'$ we assume that these signs are equal (in this case there is no sign cancellation in Eq. \ref{bprimed}). Then we get the following limits:
\begin{equation}
\label{constraints}
|b'|<0.26 \quad \quad {(\rm 95 \% \; CL)}; \quad 
|b'|<0.31 \quad \quad {(\rm 99.7 \% \; CL)}. 
 \end{equation}
 
\section{Right-handed couplings and neutrino mass determination in KATRIN}
\medskip

\begin{figure}[h]
\begin{minipage}[b]{0.80\textwidth}
\raisebox{-1cm}{\includegraphics[width=0.80\textwidth]{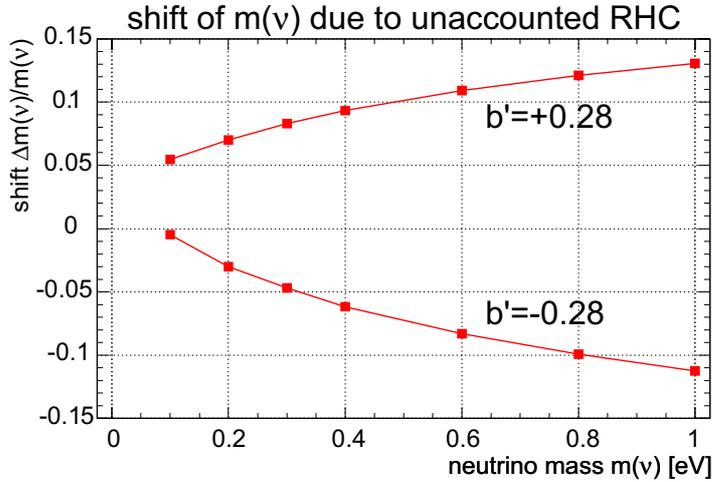}}
\end{minipage}
\begin{minipage}[b]{0.19\textwidth}
\caption{\label{fig2} Relative shift 
$(m_\nu^{\rm (fit)}-m_\nu^{\rm (real)})/m_\nu^{\rm (real)}$
of neutrino mass due to unaccounted right-handed couplings, as
function of $m_\nu^{\rm (real)}$.}
\end{minipage}
\end{figure}

Let us assume that the real value of the parameter $b'$ is nonzero, and the KATRIN data are analyzed with $b'=0$ theory
(Standard Model). In this case, the fitted neutrino mass value should deviate from the real mass value. 
Fig. \ref{fig2} shows the $\Delta m_\nu
/m_\nu=(m_\nu^{\rm (fit)}-m_\nu^{\rm (real)})/m_\nu^{\rm (real)}$ 
relative deviation due to the unaccounted right-handed  
parameter $b'=\pm 0.28$.  
The KATRIN design parameters and the statistical method described in
Ref. \cite{DesignReport2004} have been used for this calculation.
The fitted parameter in these calculations is the neutrino mass squared,
not the mass. 
One has to emphasize also that the endpoint was taken as a free parameter. According to
Fig. \ref{fig2} {\bf the relative change of the neutrino mass due to the unaccounted right-handed couplings is of order of 5-10 \%.} For small neutrino mass values (below 0.5 eV) the shift
$m_\nu^{\rm (fit)}-m_\nu^{\rm (real)}$ 
is smaller than the expected experimental error of the mass, for larger mass values (above 0.5 eV) the shift of the mass is larger than the experimental error.

Taking the endpoint as a fixed input parameter, the results are completely
different. To illustrate this difference, let us consider a special numerical example: we assume that the real neutrino mass is
$m_\nu^{\rm (real)}$=0.35 eV, and the real value of the parameter $b'$ is
$b'_{\rm real}=\pm 0.28$. Then we make a computer experiment: we generate the KATRIN data by using these real values, but we analyze the data assuming $b'=0$. Table \ref{table1} shows the fitted neutrino mass values
of these calculations with fixed and with free endpoint. {\bf With free endpoint the fitted mass values are close to the real mass. On the other hand, in the case of fixed endpoint
the fitted neutrino mass with $b'_{\rm real}=-0.28$ is completely different from the real mass value. In the case of $b'_{\rm real}=+0.28$
the fitted mass squared becomes negative, in spite of the positive
real mass value. Using the endpoint as a free parameter such a large deviation between real and fitted mass or mass squared values does not occur.} 

\begin{table}[h]
\begin{tabular}{lll}
\br
$b'_{\rm real}$ & $E_0$ fixed & $E_0$ free \\
\mr
-0.28 &   $m_\nu^{\rm (fit)}$=0.6 eV & $m_\nu^{\rm (fit)}$=0.33 eV \\
+0.28 &  $m_\nu^{2\;{\rm (fit)}} $=-0.1 eV$^2$  & $m_\nu^{\rm (fit)}$=0.38 eV\\
\br
\end{tabular}
\hspace{0.06\textwidth}
\begin{minipage}[c]{0.4\textwidth}
\caption{\label{table1} Fitted neutrino mass (or mass squared) values
with $m_\nu^{\rm (real)}$=0.35 eV. }
\end{minipage}
\end{table}

Several theoretical publications present large right-handed coupling effects to the neutrino mass determination (Refs. \cite{Stephenson98,Stephenson00,Ignatiev}). Refs. \cite{ Stephenson98,Stephenson00} tried to explain the negative mass squared anomaly of several neutrino mass experiments by assuming the presence of non-zero right-handed couplings. {\bf Nevertheless, all these 3 publications used in their analyses fixed endpoint, therefore they are not relevant for
the neutrino mass experiments (like KATRIN) using free endpoint.} 
We mention that in Ref. 
\cite{Kraus} right-handed couplings were searched in the data of the Mainz neutrino mass experiment, using free endpoint in the analysis; the data did not favor the existence of non-zero right-handed couplings.

\end{document}